\documentclass[aps, prb, twocolumn, amsmath,amssymb,showpacs,superscriptaddress]{revtex4-1}

\usepackage{graphicx}
\usepackage{color}
\usepackage{amsthm}
\usepackage{amsmath}
\usepackage{amssymb}
\usepackage{setspace}
\usepackage{braket}
\usepackage{float}
\usepackage{subfig}
\usepackage{natbib}

\makeatletter
\@ifundefined{textcolor}{}
{
 \definecolor{BLACK}{gray}{0}
 \definecolor{WHITE}{gray}{1}
 \definecolor{RED}{rgb}{1,0,0}
 \definecolor{GREEN}{rgb}{0,1,0}
 \definecolor{BLUE}{rgb}{0,0,1}
  \definecolor{purple}{rgb}{0.5,0,1}
 \definecolor{CYAN}{cmyk}{1,0,0,0}
 \definecolor{MAGENTA}{cmyk}{0,1,0,0}
 \definecolor{YELLOW}{cmyk}{0,0,1,0}
}
\newcommand{\E}{\mathcal{E}}

\newcommand{\bk}{\mathbf{k}}

\newcommand{\kb}{k_{B}}

\begin{document}

\title{Nernst Thermopower of Time-Reversal Breaking Type-II Weyl Semimetals}

\author{Robert C. McKay}
\email[]{rcmckay2@illinois.edu}
\affiliation{Department of Physics, University of Illinois Urbana-Champaign, Urbana, IL 61801, USA}

\author{Timothy M. McCormick}
\affiliation{Department of Physics and Center for Emergent Materials, The Ohio State University, Columbus, OH 43210, USA}

\author{Nandini Trivedi}
\email[]{trivedi.15@osu.edu}
\affiliation{Department of Physics and Center for Emergent Materials, The Ohio State University, Columbus, OH 43210, USA}

\date{\today}

\begin{abstract}
Weyl semimetals host linear energy dispersions around Weyl nodes, as well as monopoles of Berry curvature in momentum space around these points.  These features give rise to unique transport signatures in a Weyl semimetal, such as transverse transport without an applied magnetic field, known as anomalous transport.  The type-II Weyl semimetal, recently experimentally demonstrated in several materials, is classified by a tilting of the Weyl nodes.  This paper provides a theoretical study on thermoelectric transport in time-reversal breaking type-II Weyl semimetals.  Our results examine the balance between anomalous and non-anomalous contributions to the Nernst effect when subject to an external magnetic field.  We also show how increasing scattering times have on enhancing effect on thermoelectric transport in these materials.  Since a temperature-dependent chemical potential has been theoretically shown to be paramount when considering anomalous transport, we also study how similar considerations impact the Nernst thermopower in the non-anomalous case.
\end{abstract}
\maketitle

\section{Introduction}

Weyl semimetals have generated exceptional attention in recent years \cite{felserReview, hasanReview, vishReview}.  Weyl semimetals are characterized by their linear energy dispersion with massless excitations, known as Weyl fermions \cite{Weyl1929}.  These linear disperions meet at distinct points, called Weyl nodes, which come in pairs.  Another prominent feature of Weyl semimetals is Berry curvature, which acts like a magnetic field monopole in momentum space around each of the nodal pairs \cite{FangMonopole}.  Pairs of Weyl nodes will come in opposite chirality, corresponding to Berry monopoles of positive and negative charge \cite{Nielsen}.  The presense of topologically protected Fermi arcs, open contours of states on the Brillouin zone surface, is a unique consequence of bulk Berry curvature in Weyl semimetals \cite{wanTurnVish}.  Bulk Weyl fermions and Fermi arcs are key features of a Weyl semimetal's characteristic behavior and allow for experimentally accessible attributes.
\par These unique characteristics have been probed in various newly-discovered Weyl semimetals in recent years.  TaAs is the first of these materials, whose Weyl-characteristic Fermi arc was reported in ARPES \cite{Lv2015, LvPhyRevX2015, Xu613}.  Additionally, type-II Weyl semimetals have also recently been predicted and confirmed.  A type-II Weyl semimetal differs from a type-I Weyl semimetal in that type-II contains tiled Weyl nodes whereas type-I contains Weyl nodes that are perpendicular to its momentum plane \cite{PhysRevB.78.045415, Soluyanov2015, PhysRevLett.117.056805, Sun2015}.  The transition metal dichalcogenide $\text{MoTe}_\text{2}$ has shown to be a type-II Weyl semimetal, both from ARPES data as well as DFT \cite{PhysRevLett.117.056805, Huang2016, JiangMoTe2}.  A key attribute of type-II Weyl semimetals is that their tilted Weyl nodes lead to a finite density of electrons and holes at the Weyl energy, where the nodal points exist. 

Another important feature of Weyl semimetals is the behavior of transport in these materials.  The Berry monopoles generate distinct signatures for a Weyl semimetal.  For example, Weyl semimetals will exhibit a chiral anomaly when an electric field and magnetic field are applied in parallel directions \cite{PhysRevB.89.195137, PhysRevB.88.104412, PhysRevB.92.075205, PhysRevB.89.085126, Zhang2016, Hirschberger2016}.  The chiral anomaly generates anomalous transport and negative longitudinal magnetoresistance, which have been experimentally observed \cite{PhysRevX.5.031023, Du59:657406}.  Furthermore, both theoretical and experimental results for thermoelectric transport have come out in recent years \cite{nbpNernst, fieteThermoelec, PhysRevB.93.035116, arcTherm}.

Previous work in anomalous transport without an applied external magnetic field in type-II Weyl semimetals have indicated that the tilt in a type-II Weyl semimetal has broad impacts on the electric, thermal, and thermoelectric transport properties as well as the Nernst effect \cite{mccormickmckay, wte2liftrans, NernstFerreiros}.  Additionally, it has been shown that the role of the  temperature dependent chemical potential yields important contributions when considering anomalous transport.  This paper extends these results to include non-anomalous contributions to the transport, and investigate similar phenomena in the Nernst thermopower.

A primary purpose of this paper is to include the magnetic field when examining the Nernst effect in type-II Weyl semimetals.  This paper differs from its predecessors in that, not only does it include a magnetic field and temperature-dependent chemical potential, it also encompasses a scattering time.  While preliminary research has been done in electrical conductivity, thermal conductivity, and thermoelectric conductivity for Weyl semimetals \cite{mccormickmckay, fieteThermoelec, PhysRevB.93.035116}, we seek to expand upon these ideas with the Nernst effect.  The Nernst effect is more experimentally pertinent, such as in the case of the Weyl semimetal NbP \cite{nbpNernst}.  The model used in this paper strives to capture Nernst thermopower from the minimal working lattice model for a time-reversal breaking type-II Weyl semimetal \cite{mkt}.

In this paper, Section II  discusses the model we use along with calculations involving Berry curvature, temperature dependence on the chemical potential, and semi-classical Boltzmann transport.  Section III shows the results of our transport calculations for a variety of temperatures, magnetic fields, scattering times, Weyl node tilts, and Fermi energies.  We also look at the effect of applying a magnetic field and the effect of a temperature-dependent chemical potential.  Section IV presents remaining conclusions, future outlooks, and closing remarks.

\section{Model}

\begin{figure*}
    \centering
	\includegraphics
	[width=0.7\textwidth]
	{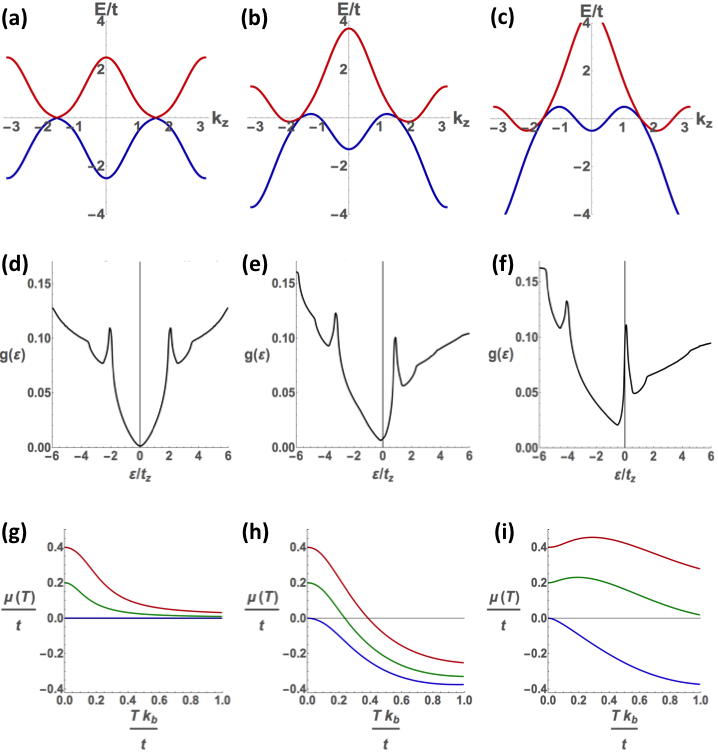}
	\caption {The parameters $m = 3t$, $t_z = t$, and $k_0 a = \pi/2$ were used in each of the above plots.  (a-c) shows the energy dispersion from the Hamiltonian listed in Equation \eqref{ham}.  The electron band is in red while the hole band is in blue.  Columns (a), (d), and (g) are at $\gamma=0$; Columns (b), (e), and (h) are at $\gamma=1.2 t$; and columns (c), (f), and (j) are at $\gamma=2 t$.  Plots (d-f) show the density of states, $g(\E)$, with the local minimum shifting with $\gamma$.  The plots (g-i) are of the chemical potential, $\mu(T)$, at various values of the Fermi energy: $\E_F=0.4 t$ is red, $\E_F=0.2 t$ is green, and $\E_F=0$ is blue.}
	\label{Figure1}
\end{figure*}

We use the following Hamiltonian model for a time-reversal breaking Weyl semimetal\cite{mkt}:
\begin{multline}
\label{ham}
\hat{H} = \gamma(\cos(k_z a) - \cos(k_0 a)) \hat{\sigma}_{0} - 2t \sin(k_x a) \hat{\sigma}_{1}\\ - 2t \sin(k_y a) \hat{\sigma}_{2}  - \Big[ 2 t_{z} \big( \cos(k_{z} a) - \cos(k_{0} a) \big) \\+ m (2 - \cos(k_{x} a) - \cos(k_y a))  + \gamma_z (\cos(3 k_{z} a) - \cos(3 k_{0} a)) \Big]\hat{\sigma}_{3},
\end{multline}
where $\hat{\sigma}_0$ is the identity matrix, $a$ is the lattice spacing, $t$, $t_z$, $\gamma_z$, and $m$ are all energy hopping parameters, $k_{0}$ is the distance between Weyl nodes in $k$-space, and $\gamma$ controls the tilt of the Weyl cones.  The energy, $\E(k_x, k_y, k_z)$, that results from Equation \eqref{ham} for several $\gamma$ values is shown for $k_x=k_y=0$ in Fig. \ref{Figure1}(a-c).

\par We chose this Hamiltonian over the continuum Hamiltonian.  The continuum model is given as
\begin{equation}
\label{contham}
\hat{H}_W = \gamma k_z \hat{\sigma}_0 +\chi \hbar v_F \big(\bk - \mathbf{k}_0 \big) \cdot \mathbf{\sigma}, 
\end{equation}
where $\chi$ is chirality, $v_F$ is the Fermi speed, and $\mathbf{k}_0$ is the spacing between Weyl nodes.  Since the type-I to type-II transition occurs at $\gamma = \hbar v_F$, the continuum model is useful for $\gamma<\hbar v_F$.  However, in the type-II regime, the continuum model in Equation \eqref{contham} is unphysical.  At $\gamma > \hbar v_F$, Equation \eqref{contham} produces an electron and hole pocket that are not closed for larger values of $\mathbf{k}$ while Equation \eqref{ham} yields Fermi pockets that are bounded in these regions.  This discrepancy in models prompted us to use Equation \eqref{ham} when considering transport.  Equation \eqref{ham} also allows for a separation in pairs of electron and hole pockets centered around each of the Weyl nodes.  This type-I to type-II transition occurs at $\gamma=2 t_z - 3 \gamma_z$.

\subsection{Berry Curvature}

Equation \eqref{ham} can be written in the form $\hat{H} = d_{0}(\bk) \hat{\sigma}_{0} + \mathbf{d}(\bk) \cdot \mathbf{\sigma}$.  The purpose of expressing Equation \eqref{ham} in this way is to calculate the Berry curvature, which is defined by\cite{bernBook},
\begin{equation}
\label{berryDef}
\Omega_{n,i}(\bk) = \epsilon_{ijl} (-1)^{n} \dfrac{\mathbf{d}\cdot \left( \partial_{k_j} \mathbf{d} \times \partial_{k_l} \mathbf{d} \right)}{2|\mathbf{d}|^3}.
\end{equation}
$\epsilon_{ijl}$ is the Levi-Civita tensor and $n$ denotes the $n$-th band.  Berry curvature will play an important role in transport, taking a role similar to a magnetic field.  Although $\gamma$ will influence the behavior of transport, it will not impact the Berry curvature since $d_0(\mathbf{k})$ does not enter into Equation \eqref{berryDef}.  Another feature that Berry curvature produces in Weyl semimetals is anomalous transport, or transverse transport without an applied magnetic field.  Berry curvature will provide a transverse contribution when considering transport, leading to anomalous transport.

\subsection{Chemical Potential's Dependence on Temperature}

Chemical potential is nearly constant with metals of high densities of electrons at degenerate temperatures.  However, the case we consider is with low density semimetals, which have strong temperature dependencies at experimentally pertinent temperatures\cite{nbpNernst,wte2liftrans}.  We can calculate the temperature dependence on the chemical potential by self-consistently solving for $\mu(T)$ for a fixed density:
\begin{equation}
\label{muoft}
n = \int_{-\infty}^{\infty} d\E \dfrac{g(\E)}{1+e^{\frac{\E-\mu(T)}{\kb T}}},
\end{equation}
where the energy $\E$ is given through the Weyl Hamiltonian from Equation \eqref{ham}, $n$ is the density, $T$ is the temperature, and $g(\E)$ is the density of states.  The density of states is given by
\begin{equation}
\label{doseqn}
g(\E) = -\dfrac{1}{\pi} \sum_n \textrm{Im}\Bigg[
\int \dfrac{d^3 k}{(2\pi)^3} G_{n}(\bk,\E)
\Bigg],
\end{equation}
where $G_{n}(\bk,\E)$ is the $n$-th band Green function.

\par In the type-I region of $\gamma$, $g(\E)$ will generically occur at the Weyl point energy due to symmetry in the electron and hole bands\cite{nbpNernst}.  This results in chemical potential shifting in the type-I Weyl semimetal around the Weyl points as temperature increases.  $\mu(T)$ will asymptotically approach the Weyl energy but will never cross this enegy level in the type-I limit.  However, in the type-II region of $\gamma$, the electron and hole bands are generally no longer symmetric about the Weyl energy as tilt becomes more manifest.  This asymmetry between the bands leads to the minimum value shifting in $g(\E)$ above or below the Weyl energy.  Example plots of $g(\E)$ for different values of $\gamma$ are shown in Fig. \ref{Figure1}(d-f).  At $\gamma=0$, we see $g(\E)$ achieves its minimum value at the nodal energy, $\E=0$.  For values of $\gamma$ in the type-II region, we see the nodal tilt breaks the electron-hole symmetry, resulting in a shift in the minimum of $g(\E)$ away from the nodal energy.  Fig. \ref{Figure1}(d-f) illustrates the effect of this shifting minimum of $g(\E)$ on the chemical potential for a variety of $\gamma$'s.  For low values of $\gamma$, Fig. \ref{Figure1}(h) and Fig. \ref{Figure1}(i) shows that the chemical potential shifts to the Weyl energy.  The $\mu(T)$ curves shift to where the Weyl energy occurs on a scale that is roughly the distance that $\E_F$ is away from the energy that $g(\E)$ achieves its minimum.  For larger values of $\gamma$, the temperature scale over which $\mu(T)$ shifts to the Weyl node energy becomes larger than the appropriate scales we consider.

\subsection{Transport Definition}

The linear responses to an electric field $\mathbf{E}$ and temperature gradient $-\nabla T$ are given through the Onsager transport equations\cite{harmHonig}:
\begin{equation}
\label{onsagerDef}
\left( \begin{array}{c}
\mathbf{J}^e \\
\mathbf{J}^q 
\end{array} \right)
=
\left( \begin{array}{cc}
\mathbf{L}^{EE} & \mathbf{L}^{ET} \\
\mathbf{L}^{TE} & \mathbf{L}^{TT}
\end{array} \right)
\cdot 
\left( \begin{array}{c}
\mathbf{E} \\
-\nabla T 
\end{array} \right).
\end{equation}
$\mathbf{J}^e$ is the electric current density and $\mathbf{J}^q$ is the heat current density.  The transport coefficients in Eqn. \eqref{onsagerDef} can be calculated by solving the non-equilibrium distribution function with the Boltzmann formalism.  For a scattering time $\tau$, we used the relaxation time approximation, $-\frac{f - f_{0}}{\tau}$, in the steady-state Boltzmann equation so that the non-equilibrium distribution, $f$, is given by\cite{PhysRevB.93.035116,fieteThermoelec,ashcroftbook, doubleWeyl}, 
\begin{equation}
\label{BoltzmannFlowEq}
(\frac{\partial}{\partial t}+\mathbf{\dot{r}} \cdot \nabla_{\mathbf{r}} + \mathbf{\dot{k}} \cdot \nabla_{\mathbf{k}})f = -\frac{f - f_{0}}{\tau}.
\end{equation}

The equations for electric current density, $\mathbf{J}^e$ and heat current density $\mathbf{J}^q$ are respectively given as\cite{ashcroftbook, anomNernstNiu, thermhall3,fieteThermoelec,mccormickmckay, doubleWeyl},
\begin{widetext}
\begin{align}
\label{BoltzmannJeEq}
\begin{split}
\mathbf{J}^e =  -e\int \frac{d^3 \mathbf{k}}{(2 \pi)^3}(\mathbf{v} + \frac{e}{\hbar}\mathbf{E} \times \mathbf{\Omega})f  + \frac{\nabla T}{T} \times (\frac{e}{\hbar}\int \frac{d^3 \mathbf{k}}{(2 \pi)^3} \mathbf{\Omega} [(\E-\mu)f_{0} + k_{B}T \ln{(1 + e^{-\beta (\E - \mu)})}])
\end{split}
\end{align}
and
\begin{align}
\label{BoltzmannJqEq}
\begin{split}
\mathbf{J}^q =&  \int \frac{d^3 \mathbf{k}}{(2 \pi)^3} (\E - \mu) \mathbf{v} f 
+ \int \frac{d^3 \mathbf{k}}{(2 \pi)^3} (\mathbf{E} \times \frac{e}{\hbar} \mathbf{\Omega} \{(\E-\mu)f_{0} + k_{B}T \ln{1 + e^{-\beta (\E - \mu)}}\})
\\ & 
+ \frac{e k_{B} \nabla T}{\beta \hbar} \times \int \frac{d^3 \mathbf{k}}{(2 \pi)^3} \mathbf{\Omega} [\frac{\pi^2}{3}+f_0 \ln^2{(1/f_0-1)}  - \ln^2{(1-f_0)} -2\text{Li}_{2}(1-f_0)].
\end{split}
\end{align}
\end{widetext}
The group velocity is $\mathbf{v} \equiv \nabla_{\mathbf{k}}\E$, $\beta = \frac{1}{k_{B} T}$ for temperature $T$, $f_{0}$ is the equilibrium Fermi-Dirac distribution, $\mathbf{\Omega}$ is the Berry field, and $\text{Li}_{n}(z) = \sum^{\infty}_{k=1}\frac{z^k}{k^n}$ is the Jonquière polylogarithmic function.
\par With the current densities defined in Equations \eqref{BoltzmannJeEq} and \eqref{BoltzmannJqEq}, Equation \eqref{onsagerDef} can be used to solve for individual transport coefficients.  With an applied magnetic field $\mathbf{B} = B \mathbf{e}_z$ along the $z$-direction, the non-equilibrium distribution can be solved through Equation \eqref{BoltzmannFlowEq} and the two equations of motion for a particle of charge $q$ in the $n$-th band: $\mathbf{\dot{r}}_{n} = \frac{1}{\hbar}\mathbf{\nabla}_\bk \E_{n}(\bk) - \big(\mathbf{\dot{k}} \times \mathbf{\Omega}_{n}(\bk) \big)$ and $\mathbf{\dot{k}}=q \mathbf{E} + q \mathbf{\dot{r}}_{n} \times \mathbf{B}$.  The $\mathbf{\dot{k}} \times \mathbf{\Omega}_{n}(\bk)$ term is the anomalous velocity term, arising from the Berry field.  These series of equations describes the electric transport coefficients and thermoelectric transport coefficients.  In solving for these coefficients, it is useful to define the following relations
\cite{PhysRevB.93.035116}:
\begin{widetext}
\begin{equation}
\label{cx}
c_{x}=e B D \frac{[\frac{v_x}{m_{xy}}-\frac{v_y}{m_{xx}}][-\frac{e B v_y}{m_{xx}}+ \frac{e B v_x}{m_{xy}}-\frac{v_x}{D \tau}] + [\frac{v_x}{m_{yy}}+\frac{v_y}{m_{xy}}][-\frac{e B v_y}{m_{xy}}+ \frac{e B v_x}{m_{yy}}-\frac{v_y}{D \tau}]}{[-\frac{e B v_y}{m_{xx}} + \frac{e B v_x}{m_{xy}} - \frac{v_x}{D \tau}]^2 + [-\frac{e B v_y}{m_{xy}} + \frac{e B v_x}{m_{yy}} - \frac{v_y}{D \tau}]^2}
\end{equation}
and
\begin{equation}
\label{cy}
c_{y}=e B D \frac{[\frac{v_x}{m_{xy}}-\frac{v_y}{m_{xx}}][-\frac{e B v_y}{m_{xy}}+ \frac{e B v_x}{m_{yy}}-\frac{v_y}{D \tau}] - [\frac{v_x}{m_{yy}}+\frac{v_y}{m_{xy}}][-\frac{e B v_y}{m_{xx}}+ \frac{e B v_x}{m_{xy}}-\frac{v_x}{D \tau}]}{[-\frac{e B v_y}{m_{xx}} + \frac{e B v_x}{m_{xy}} - \frac{v_x}{D \tau}]^2 + [-\frac{e B v_y}{m_{xy}} + \frac{e B v_x}{m_{yy}} - \frac{v_y}{D \tau}]^2},
\end{equation}
\end{widetext}
where $D = (1+\frac{e}{\hbar}\mathbf{B} \cdot \mathbf{\Omega})^{-1}$ is a coupling term of the magnetic field and the Berry field\cite{fieteThermoelec}.  Additionally, $m_{ij}$ is the $ij$-th entry of the effective mass tensor, where the inverse effective mass is defined as $m_{ij}^{-1}=\frac{1}{\hbar^2} (\frac{\partial^2}{\partial k_i \partial k_j}\E)$\cite{lauer}.

\par With the definitions of Equations \eqref{cx} and \eqref{cy}, we can define the conductivity coefficients:
\begin{equation}
\label{LEExx}
L^{EE}_{xx}=\frac{e^2}{\hbar}\int \frac{d^3 \mathbf{k}}{(2 \pi)^3}v_{x}^2 \tau (-\frac{\partial f_0}{\partial \epsilon})(c_x - D)
\end{equation}
\begin{equation}
\label{LEExy}
\begin{split}
L^{EE}_{xy}=&\frac{e^2}{\hbar}\int \frac{d^3 \mathbf{k}}{(2 \pi)^3}[v_{y}^2 c_y + v_x v_y (c_x - D)]\tau (-\frac{\partial f_0}{\partial \epsilon}) \\& + \frac{e^2}{\hbar} \int \frac{d^3 \mathbf{k}}{(2 \pi)^3} \Omega_z f_{0}
\end{split},
\end{equation}
and the thermoelectric transport coefficients:
\begin{equation}
\label{LETxx}
L^{ET}_{xx}=\frac{k_{B} e}{\hbar} \int \frac{d^3 \mathbf{k}}{(2 \pi)^3}v_{x}^2 \tau \frac{\epsilon - \mu}{T}(-\frac{\partial f_0}{\partial \epsilon})(c_x - D)
\end{equation}

\begin{equation}
\label{LETxy}
\begin{split}
L^{ET}_{xy}=&\frac{k_{B} e}{\hbar} \int \frac{d^3 \mathbf{k}}{(2 \pi)^3}[v_{y}^2 c_y + v_x v_y (c_x - D)]\tau \frac{\epsilon - \mu}{T} (-\frac{\partial f_0}{\partial \epsilon}) \\ &+ \frac{k_{B} e}{\hbar} \int \frac{d^3 \mathbf{k}}{(2 \pi)^3} \Omega_z s_{k},
\end{split}
\end{equation}

\begin{figure*}
    \centering
	\includegraphics[width=1\textwidth]
	{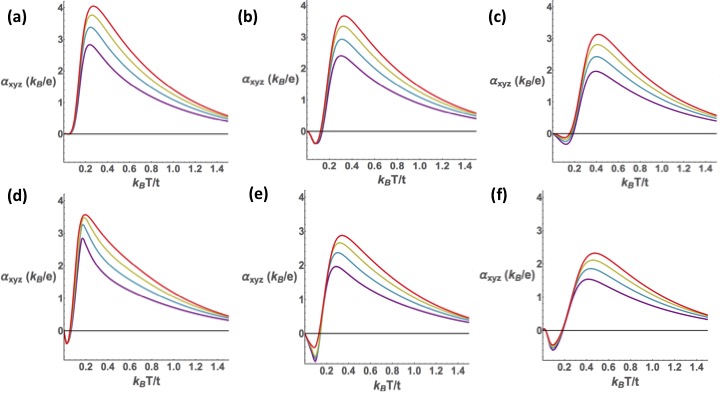}
	\caption{Nernst thermopower as a function of temperature for Eqn. (\ref{ham}) with parameters $m = 3t$, $t_z = t$, $k_0 a = \pi/2$, and $B=0.0015 e a^2/\hbar$.  Row (a-c) is at $\gamma=0$ and row (d-f) is at $\gamma=1.2 t$. In terms of the Fermi energy: column (a) and (d) are at $\E_F=0$, column (b) and (e) are at $\E_F=0.2 t$, and column (c) and (f) are at $\E_F=0.4 t$.  The colors are purple at $\tau=150 \hbar/t$, blue at $\tau=200 \hbar/t$, yellow at $\tau=150 \hbar/t$, and red at $\tau=300 \hbar/t$.}
	\label{Figure2}
\end{figure*}

where $s_{k}=-f_{0} \ln{f_{0}} - (1 - f_{0}) \ln(1 - f_{0})$ is the entropy density function.  This also describes the anomalous contributions to the transverse transport coefficients, taking $B=0$, which becomes\cite{mccormickmckay}
\begin{equation}
\label{anomLEE}
L^{EE}_{xy\text{, anomalous}} = \frac{e^2}{\hbar} \int \frac{d^3 \mathbf{k}}{(2 \pi)^3} \Omega_z f_{0}
\end{equation}
and
\begin{equation}
\label{anomLET}
L^{ET}_{xy\text{, anomalous}} = \frac{k_{B} e}{\hbar} \int \frac{d^3 \mathbf{k}}{(2 \pi)^3} \Omega_z s_{k}.
\end{equation}

\par We will examine the experimentally relevant quantity, the Nernst effect\cite{nbpNernst}, which is described in terms of the Onsager transport coefficients: 
\begin{equation}
 \label{NernstEq}
\alpha_{xyz} = \dfrac{E_y}{-\nabla_x T}=\dfrac{L^{EE}_{xx} L^{ET}_{xy} - L^{EE}_{xy} L^{ET}_{xx}}{(L^{EE}_{xx})^2+(L^{EE}_{xy})^2}.
\end{equation}
The notation of $\alpha_{xyz}$ indicates an electric field responding in the $\hat{x}$ direction, a temperature gradiant in the $\hat{y}$ direction, and an applied magnetic field in the $\hat{z}$ direction.

\section{Transport Results}

To calculate the Nernst effect in the type-I and type-II Weyl systems, we put $\mathbf{B}$ in the $\hat{z}$ direction and a $-\nabla T$ in the $\hat{y}$ direction.  
\subsection{Scattering dependence of Nernst effect}

\subsubsection{Nernst Effect vs. Temperature}

\par We will first consider the Nernst effect as functions of temperature for various $\tau$ and Fermi energy values shown in in Figure \ref{Figure2}.  As $\tau$ increases, the Nernst effect is enhanced as a function of temperature, which is shown in Figure \ref{Figure1}.  The maximas of the Nernst thermopower generally increase in magnitude and occur at higher temperatures as $\tau$ increases.  The peak's tendency toward positive temperatures with increasing $\tau$ is due to the $c_x$ in the $(c_x - D)$ terms in Equations \eqref{LETxx} and \eqref{LETxy}. In Equation \eqref{cx} and Equation \eqref{cy} for large values of $\tau$, $c_y$ and $c_x$ both go as  $\sim \tau$ to lowest order.  This leads to an overall positive shift in the integrands for the thermoelectric coefficients.  
\begin{figure*}
    \centering
	\includegraphics[width=1\textwidth]
	{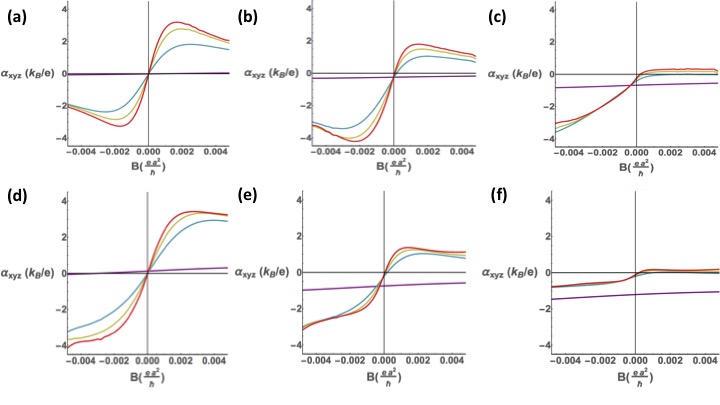}
	\caption{The parameters $m = 3t$, $t_z = t$, $k_0 a = \pi/2$, and $T k_B=0.201 t$ were used in the above plots, showing the Nernst effect as a function of the magnetic field strength.  Row (a-c) is at $\gamma=0$ and row (d-f) is at $\gamma=1.2 t$. In terms of the Fermi energy: column (a) and (d) are at $\E_F=0$, column (b) and (e) are at $\E_F=0.2 t$, and column (c) and (f) are at $\E_F=0.4 t$.  Purple $\tau=10 \hbar/t$, blue $\tau=100 \hbar/t$, yellow $\tau=150 \hbar/t$, and red $\tau=200 \hbar/t$.}
	\label{Figure3}
\end{figure*}

The increase in the apex values of $\alpha_{xyz}$ is explained by the scattering times multiplying the entirety of Equations \eqref{LETxx} to \eqref{LETxy}.
Each of the transport coefficients have a $\tau$ multiplying their integrals so that as $\tau$ increases, so does the overall magnitude of the transport coefficients.  However, the Nernst effect, from Equation \eqref{NernstEq}, relies on $L^{ET}_{xy}$ and $L^{ET}_{xx}$ in the numerator, which each go as $\sim 1/T$.  The result is the most prominent contributions to the thermoelectric coefficient are when the temperature is lower and tapers off with higher temperatures.

\subsubsection{Nernst Effect vs. Magnetic Field Strength}

\par We now examine the Nernst effect as a function of magnetic field strength for numerous scattering times, shown in Figure \ref{Figure2}.  Two distinct regions are present.  One region is the low scattering limit, where the field dependence of the Nernst thermopower is close to linear; the other is the high scattering limit, where the Nernst thermopower is nearly antisymmetric with respect to the field.  The slight asymmetry is due to the time-reversal breaking and is a result of the anomalous effects from Berry curvature, so that at $B=0$ $\alpha_{xyz}$ is non-zero.  This offset contribution is given in Equations \eqref{anomLEE} and \eqref{anomLET}.  Although most apparent at $B=0$, the anomalous contributions permeate for all values of $B$.  Since the time-reversal symmetry is broken in the Hamiltonian from Equation \eqref{ham}, this anomalous offset from pure antisymmetry must always remain\cite{mccormickmckay}.
\par In Figure \ref{Figure2} and \ref{Figure3}, we see that as we increase the Fermi energy, $\E_F$, from left to right in the figure, that curve shapes are mostly unaffected.  The main difference is the scaling of Figure \ref{Figure3}, which compresses and shifts toward $-\alpha_{xyz}$.  Since $f_0$ goes as $e^{\frac{\E-\mu}{k_B T}}$ and since $\mu(T)$ will generally experience an upward shift as $\E_F$ increases as shown in Figure \ref{Figure1}(g-i), then it follows that as $\mu(T)$ increases from $\E_F$, then $f_0$ will both broaden and shift with $T$.  Since $f_0$ is spreading out over a broader range of $T$, then $\frac{f- f_0}{\tau}$ from Equation \eqref{BoltzmannFlowEq} will be more sensitive over a larger range of temperatures.  The manifestation of this sensitivity is that a change in $\E_F$ generates a change in the $\alpha_{xyz}$ for small values of $\tau$.

\subsection{Nernst Effect for a Variety of Temperatures and Magnetic Fields}

\begin{figure*}
    \centering
	\includegraphics[width=1\textwidth]
	{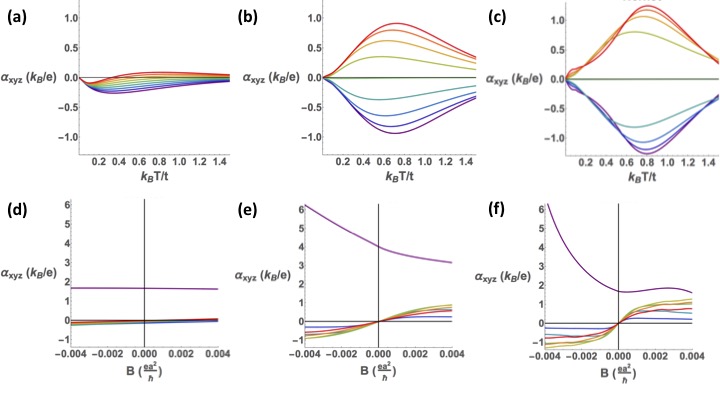}
	\caption{Nernst effect for Eqn. (\ref{ham}) with parameters $m = 3t$, $t_z = t$, $k_0 a = \pi/2$, $\gamma=2 t$, and $\E_F=0.2 t$.  Row (a-c) shows $\alpha_{xyz}$ vs. $T$ and row (d-f) shows $\alpha_{xyz}$ vs. $B$. In terms of the scattering time: column (a) and (d) are at $\tau=10 \hbar/t$, column (b) and (e) are at $\tau =100 \hbar/t$, and column (c) and (f) are at $\tau=300 \hbar/t$.  The color scheme for (a-c) represents: purple as $B=-0.004 e a^2/\hbar$, dark blue as $B=-0.003 e a^2/\hbar$, blue as $B=-0.002 e a^2/\hbar$, teal as $B=-0.001 e a^2/\hbar$, green as $B=0$, yellow as $B=0.001 e a^2/\hbar$, light orange as $B=0.002 e a^2/\hbar$, dark orange as $B=0.003 e a^2/\hbar$, and red as $B=0.004 e a^2/\hbar$.  For plots (d-f) the color scheme indicates: purple as $T=0.001 t/k_B$, dark blue as $T=0.201 t/k_B$, teal as $T=0.401 t/k_B$, green as $T=0.601 t/k_B$, gold as $T=0.801 t/k_B$, orange as $T=1.001 t/k_B$, and red as $T=1.201 t/k_B$}.
	\label{Figure4}
\end{figure*}

\subsubsection{Nersnt Effect vs. Temperature}

\par Plots of the Nernst effect as a function of the temperature for a variety of magnetic fields are given in Figure \ref{Figure4}(a-c). The chemical potential crossing the nodal energy with temperature has an important effect on the temperature dependence of $\alpha_{xyz}$.  The energy at which $\mu(T)$ crosses the Weyl energy is also the energy at which the Berry curvature is the strongest, so $\mu(T)$ has important consequences on transport.  $c_x$ and $c_y$ from Equations \eqref{cx} and \eqref{cy} go like $\sim B$.  For $|B|$, the magnitude of the non-anomalous parts  will tend to enhance for Equations \eqref{LEExx} to \eqref{LETxy}.  However, entries in $L^{ET}$ differ from entries in $L^{EE}$ by $\frac{\E - \mu(T)}{k_B T}$.  So, as temperature changes in $\alpha_{xyz}$, the modulated magnetic field is most strongly represented in the $L^{ET}$ contribution.  As temperature increases, the effects of the Fermi distribution's temperature dependence wash out terms that go as $\frac{\E - \mu(T)}{k_B T}$.  This interplay between the Fermi distribution terms and terms that extinguish with increasing $T$ yields peaks that shift for varying magnetic field in the Nernst effect.

\par As the scattering time goes from a low value of $10 \hbar/t$(Figure \ref{Figure4}(a)), to a medium value of $100 \hbar/t$ (Figure \ref{Figure4}(b)), and a high value of $300 \hbar/t$ (Figure \ref{Figure4}(c)), the scattering time impacts the degree to which the anomalous transport is expressed from Equations \eqref{anomLEE} and Equations \eqref{anomLET}.  The Nernst curves around $\alpha_{xyz}=0$ in the medium and high scattering regimes are nearly symmetric, where the asymmetry from anomalous contributions is washed out.  Higher $\tau$ values de-emphasize the role of anomalous transport in the Nernst effect.  A slight deviation to signify the anomalous effect is barely distinct when $\alpha_{xyz}=0$ at $B=0$ in \ref{Figure4}(b).  On the other hand, at low scattering values in Figure \ref{Figure4}(a), the anomalous contribution is more apparent, with both discernible asymmetries and  $\alpha_{xyz}$ curves at nonzero field that are similarly scaled to $\alpha_{xyz}$ at $B=0$.

\subsubsection{Nersnt Effect vs. Magnetic Field Strength}

\par We now consider the Nernst effect as a function of the magnetic field strength for an assortment of temperatures in Figure \ref{Figure4}(d-f).  These curve shapes also appear antisymmetric for larger values of $\tau$ and temperatures away from $T=0$.  Terms involving $D$ and $B$ from Equations \eqref{LEExx} through \eqref{LETxy} relate the dependence here.  At smaller temperatures, the $f_0$ function from the Boltzmann equation in Equation \eqref{BoltzmannFlowEq} approaches a step function between $0$ and $1$ which depends on the sign of  $\E-\mu$.  This behavior influences how $f$ changes in the presence of a magnetic field.  Since the nonequilibrium distribution satisfies $0\leq f \leq 1$, then the left hand side of Equation \eqref{BoltzmannFlowEq} will balance this maximal difference between $f$ and $f_0$ as $f_0$ approaches a step function.  This results in $f$ becoming more sensitive to the magnetic field.  At low temperatures, the anomalous terms become prominent, as they reach a maximum value as a function of temperature (like the blue functions in Figure \ref{Figure4}(d-f)) and settle toward $0$ contribution.

\par Similar to before, the Nernst effect plots in Figure \ref{Figure4}(d-f) approach antisymmetry for larger scattering times and at large values of temperature.  Although, the deviation from antisymmetry still exists due to the anomalous contributions.  The break from antisymmetry is most evident for smaller temperatures in each of the purple curves in Figure \ref{Figure4}(d-f).  This indicates the anomalous contributions are strongest at smaller $\tau$ and $T$ values.  We also notice that as the temperature increases, the Nernst curves approach close to $\alpha_{xyz}=0$ and then rebound away, matching the previously identified behavior of the temperature dependence reaching a minimum and then decreasing.  

\section{Conclusion}

We have presented a study of the Nernst effect in type-II Weyl semimetals, where we have adjusted nodal tilt, Fermi energy, temperature, mangetic field, and scattering time.  We have demonstrated an enhancement in the Nernst effect with increasing the scattering time, and how the scattering time shifts the maximas of the Nernst curve.  We have also identified a generic reduction in the $\alpha_{xyz}$ curve with increasing Fermi energy.  Furthermore, we have found the anomalous transport behaviors are generally strongest for small scattering times and lower temperatures.  In doing all of this, we have identified the relevance of incorporating a temperature-dependent chemical potential as well.

\par We expect these results to have experimental application, similar to previous work in transport effects in NbP \cite{nbpNernst}.  We offer a comprehensive study of thermoelectric transport in time-reversal breaking Weyl semimetal.  In particular, we speculate these results are apropos to study the putative type-II time-reversal breaking Weyl semimetal $\text{YbMnBi}_\text{2}$.

\bibliography{anomTransportTypeII_ref}

\begin{thebibliography}{39}%
\makeatletter
\providecommand \@ifxundefined [1]{%
 \@ifx{#1\undefined}
}%
\providecommand \@ifnum [1]{%
 \ifnum #1\expandafter \@firstoftwo
 \else \expandafter \@secondoftwo
 \fi
}%
\providecommand \@ifx [1]{%
 \ifx #1\expandafter \@firstoftwo
 \else \expandafter \@secondoftwo
 \fi
}%
\providecommand \natexlab [1]{#1}%
\providecommand \enquote  [1]{``#1''}%
\providecommand \bibnamefont  [1]{#1}%
\providecommand \bibfnamefont [1]{#1}%
\providecommand \citenamefont [1]{#1}%
\providecommand \href@noop [0]{\@secondoftwo}%
\providecommand \href [0]{\begingroup \@sanitize@url \@href}%
\providecommand \@href[1]{\@@startlink{#1}\@@href}%
\providecommand \@@href[1]{\endgroup#1\@@endlink}%
\providecommand \@sanitize@url [0]{\catcode `\\12\catcode `\$12\catcode
  `\&12\catcode `\#12\catcode `\^12\catcode `\_12\catcode `\%12\relax}%
\providecommand \@@startlink[1]{}%
\providecommand \@@endlink[0]{}%
\providecommand \url  [0]{\begingroup\@sanitize@url \@url }%
\providecommand \@url [1]{\endgroup\@href {#1}{\urlprefix }}%
\providecommand \urlprefix  [0]{URL }%
\providecommand \Eprint [0]{\href }%
\providecommand \doibase [0]{http://dx.doi.org/}%
\providecommand \selectlanguage [0]{\@gobble}%
\providecommand \bibinfo  [0]{\@secondoftwo}%
\providecommand \bibfield  [0]{\@secondoftwo}%
\providecommand \translation [1]{[#1]}%
\providecommand \BibitemOpen [0]{}%
\providecommand \bibitemStop [0]{}%
\providecommand \bibitemNoStop [0]{.\EOS\space}%
\providecommand \EOS [0]{\spacefactor3000\relax}%
\providecommand \BibitemShut  [1]{\csname bibitem#1\endcsname}%
\let\auto@bib@innerbib\@empty
\bibitem [{\citenamefont {Yan}\ and\ \citenamefont
  {Felser}(2017)}]{felserReview}%
  \BibitemOpen
  \bibfield  {author} {\bibinfo {author} {\bibfnamefont {B.}~\bibnamefont
  {Yan}}\ and\ \bibinfo {author} {\bibfnamefont {C.}~\bibnamefont {Felser}},\
  }\href@noop {} {\bibfield  {journal} {\bibinfo  {journal} {Ann. Rev. Cond.
  Mat. Phys.}\ }\textbf {\bibinfo {volume} {8}},\ \bibinfo {pages} {337}
  (\bibinfo {year} {2017})}\BibitemShut {NoStop}%
\bibitem [{\citenamefont {Hasan}\ \emph {et~al.}(2017)\citenamefont {Hasan},
  \citenamefont {Xu}, \citenamefont {Belopolski},\ and\ \citenamefont
  {Huang}}]{hasanReview}%
  \BibitemOpen
  \bibfield  {author} {\bibinfo {author} {\bibfnamefont {M.~Z.}\ \bibnamefont
  {Hasan}}, \bibinfo {author} {\bibfnamefont {S.-Y.}\ \bibnamefont {Xu}},
  \bibinfo {author} {\bibfnamefont {I.}~\bibnamefont {Belopolski}}, \ and\
  \bibinfo {author} {\bibfnamefont {C.-M.}\ \bibnamefont {Huang}},\ }\href@noop
  {} {\bibfield  {journal} {\bibinfo  {journal} {Ann. Rev. Cond. Mat. Phys.}\
  }\textbf {\bibinfo {volume} {8}},\ \bibinfo {pages} {289} (\bibinfo {year}
  {2017})}\BibitemShut {NoStop}%
\bibitem [{\citenamefont {Armitage}\ \emph {et~al.}(2017)\citenamefont
  {Armitage}, \citenamefont {Mele},\ and\ \citenamefont
  {Vishwanath}}]{vishReview}%
  \BibitemOpen
  \bibfield  {author} {\bibinfo {author} {\bibfnamefont {N.~P.}\ \bibnamefont
  {Armitage}}, \bibinfo {author} {\bibfnamefont {E.~J.}\ \bibnamefont {Mele}},
  \ and\ \bibinfo {author} {\bibfnamefont {A.}~\bibnamefont {Vishwanath}},\
  }\href@noop {} {} (\bibinfo {year} {2017}),\ \Eprint
  {http://arxiv.org/abs/arXiv:1705.01111} {arXiv:1705.01111} \BibitemShut
  {NoStop}%
\bibitem [{\citenamefont {Weyl}(1929)}]{Weyl1929}%
  \BibitemOpen
  \bibfield  {author} {\bibinfo {author} {\bibfnamefont {H.}~\bibnamefont
  {Weyl}},\ }\href {\doibase 10.1073/pnas.15.4.323} {\bibfield  {journal}
  {\bibinfo  {journal} {Proceedings of the National Academy of Sciences of the
  United States of America}\ }\textbf {\bibinfo {volume} {15}},\ \bibinfo
  {pages} {323} (\bibinfo {year} {1929})}\BibitemShut {NoStop}%
\bibitem [{\citenamefont {Fang}\ \emph {et~al.}(2003)\citenamefont {Fang},
  \citenamefont {Nagaosa}, \citenamefont {Takahashi}, \citenamefont {Asamitsu},
  \citenamefont {Mathieu}, \citenamefont {Ogasawara}, \citenamefont {Yamada},
  \citenamefont {Kawasaki}, \citenamefont {Tokura},\ and\ \citenamefont
  {Terakura}}]{FangMonopole}%
  \BibitemOpen
  \bibfield  {author} {\bibinfo {author} {\bibfnamefont {Z.}~\bibnamefont
  {Fang}}, \bibinfo {author} {\bibfnamefont {N.}~\bibnamefont {Nagaosa}},
  \bibinfo {author} {\bibfnamefont {K.~S.}\ \bibnamefont {Takahashi}}, \bibinfo
  {author} {\bibfnamefont {A.}~\bibnamefont {Asamitsu}}, \bibinfo {author}
  {\bibfnamefont {R.}~\bibnamefont {Mathieu}}, \bibinfo {author} {\bibfnamefont
  {T.}~\bibnamefont {Ogasawara}}, \bibinfo {author} {\bibfnamefont
  {H.}~\bibnamefont {Yamada}}, \bibinfo {author} {\bibfnamefont
  {M.}~\bibnamefont {Kawasaki}}, \bibinfo {author} {\bibfnamefont
  {Y.}~\bibnamefont {Tokura}}, \ and\ \bibinfo {author} {\bibfnamefont
  {K.}~\bibnamefont {Terakura}},\ }\href {\doibase 10.1126/science.1089408}
  {\bibfield  {journal} {\bibinfo  {journal} {Science}\ }\textbf {\bibinfo
  {volume} {302}},\ \bibinfo {pages} {92} (\bibinfo {year} {2003})}\BibitemShut
  {NoStop}%
\bibitem [{\citenamefont {Nielsen}\ and\ \citenamefont
  {Ninomiya}(1981)}]{Nielsen}%
  \BibitemOpen
  \bibfield  {author} {\bibinfo {author} {\bibfnamefont {H.}~\bibnamefont
  {Nielsen}}\ and\ \bibinfo {author} {\bibfnamefont {M.}~\bibnamefont
  {Ninomiya}},\ }\href {\doibase
  http://dx.doi.org/10.1016/0370-2693(81)91026-1} {\bibfield  {journal}
  {\bibinfo  {journal} {Physics Letters B}\ }\textbf {\bibinfo {volume}
  {105}},\ \bibinfo {pages} {219 } (\bibinfo {year} {1981})}\BibitemShut
  {NoStop}%
\bibitem [{\citenamefont {Wan}\ \emph {et~al.}(2011)\citenamefont {Wan},
  \citenamefont {Turner}, \citenamefont {Vishwanath},\ and\ \citenamefont
  {Savrasov}}]{wanTurnVish}%
  \BibitemOpen
  \bibfield  {author} {\bibinfo {author} {\bibfnamefont {X.}~\bibnamefont
  {Wan}}, \bibinfo {author} {\bibfnamefont {A.~M.}\ \bibnamefont {Turner}},
  \bibinfo {author} {\bibfnamefont {A.}~\bibnamefont {Vishwanath}}, \ and\
  \bibinfo {author} {\bibfnamefont {S.~Y.}\ \bibnamefont {Savrasov}},\ }\href
  {\doibase 10.1103/PhysRevB.83.205101} {\bibfield  {journal} {\bibinfo
  {journal} {Phys. Rev. B}\ }\textbf {\bibinfo {volume} {83}},\ \bibinfo
  {pages} {205101} (\bibinfo {year} {2011})}\BibitemShut {NoStop}%
\bibitem [{\citenamefont {Lv}\ \emph {et~al.}(2015{\natexlab{a}})\citenamefont
  {Lv}, \citenamefont {Xu}, \citenamefont {Weng}, \citenamefont {Ma},
  \citenamefont {Richard}, \citenamefont {Huang}, \citenamefont {Zhao},
  \citenamefont {Chen}, \citenamefont {Matt}, \citenamefont {Bisti},
  \citenamefont {Strocov}, \citenamefont {Mesot}, \citenamefont {Fang},
  \citenamefont {Dai}, \citenamefont {Qian}, \citenamefont {Shi},\ and\
  \citenamefont {Ding}}]{Lv2015}%
  \BibitemOpen
  \bibfield  {author} {\bibinfo {author} {\bibfnamefont {B.~Q.}\ \bibnamefont
  {Lv}}, \bibinfo {author} {\bibfnamefont {N.}~\bibnamefont {Xu}}, \bibinfo
  {author} {\bibfnamefont {H.~M.}\ \bibnamefont {Weng}}, \bibinfo {author}
  {\bibfnamefont {J.~Z.}\ \bibnamefont {Ma}}, \bibinfo {author} {\bibfnamefont
  {P.}~\bibnamefont {Richard}}, \bibinfo {author} {\bibfnamefont {X.~C.}\
  \bibnamefont {Huang}}, \bibinfo {author} {\bibfnamefont {L.~X.}\ \bibnamefont
  {Zhao}}, \bibinfo {author} {\bibfnamefont {G.~F.}\ \bibnamefont {Chen}},
  \bibinfo {author} {\bibfnamefont {C.~E.}\ \bibnamefont {Matt}}, \bibinfo
  {author} {\bibfnamefont {F.}~\bibnamefont {Bisti}}, \bibinfo {author}
  {\bibfnamefont {V.~N.}\ \bibnamefont {Strocov}}, \bibinfo {author}
  {\bibfnamefont {J.}~\bibnamefont {Mesot}}, \bibinfo {author} {\bibfnamefont
  {Z.}~\bibnamefont {Fang}}, \bibinfo {author} {\bibfnamefont {X.}~\bibnamefont
  {Dai}}, \bibinfo {author} {\bibfnamefont {T.}~\bibnamefont {Qian}}, \bibinfo
  {author} {\bibfnamefont {M.}~\bibnamefont {Shi}}, \ and\ \bibinfo {author}
  {\bibfnamefont {H.}~\bibnamefont {Ding}},\ }\href
  {http://dx.doi.org/10.1038/nphys3426} {\bibfield  {journal} {\bibinfo
  {journal} {Nat Phys}\ }\textbf {\bibinfo {volume} {11}},\ \bibinfo {pages}
  {724} (\bibinfo {year} {2015}{\natexlab{a}})}\BibitemShut {NoStop}%
\bibitem [{\citenamefont {Lv}\ \emph {et~al.}(2015{\natexlab{b}})\citenamefont
  {Lv}, \citenamefont {Weng}, \citenamefont {Fu}, \citenamefont {Wang},
  \citenamefont {Miao}, \citenamefont {Ma}, \citenamefont {Richard},
  \citenamefont {Huang}, \citenamefont {Zhao}, \citenamefont {Chen},
  \citenamefont {Fang}, \citenamefont {Dai}, \citenamefont {Qian},\ and\
  \citenamefont {Ding}}]{LvPhyRevX2015}%
  \BibitemOpen
  \bibfield  {author} {\bibinfo {author} {\bibfnamefont {B.~Q.}\ \bibnamefont
  {Lv}}, \bibinfo {author} {\bibfnamefont {H.~M.}\ \bibnamefont {Weng}},
  \bibinfo {author} {\bibfnamefont {B.~B.}\ \bibnamefont {Fu}}, \bibinfo
  {author} {\bibfnamefont {X.~P.}\ \bibnamefont {Wang}}, \bibinfo {author}
  {\bibfnamefont {H.}~\bibnamefont {Miao}}, \bibinfo {author} {\bibfnamefont
  {J.}~\bibnamefont {Ma}}, \bibinfo {author} {\bibfnamefont {P.}~\bibnamefont
  {Richard}}, \bibinfo {author} {\bibfnamefont {X.~C.}\ \bibnamefont {Huang}},
  \bibinfo {author} {\bibfnamefont {L.~X.}\ \bibnamefont {Zhao}}, \bibinfo
  {author} {\bibfnamefont {G.~F.}\ \bibnamefont {Chen}}, \bibinfo {author}
  {\bibfnamefont {Z.}~\bibnamefont {Fang}}, \bibinfo {author} {\bibfnamefont
  {X.}~\bibnamefont {Dai}}, \bibinfo {author} {\bibfnamefont {T.}~\bibnamefont
  {Qian}}, \ and\ \bibinfo {author} {\bibfnamefont {H.}~\bibnamefont {Ding}},\
  }\href {\doibase 10.1103/PhysRevX.5.031013} {\bibfield  {journal} {\bibinfo
  {journal} {Phys. Rev. X}\ }\textbf {\bibinfo {volume} {5}},\ \bibinfo {pages}
  {031013} (\bibinfo {year} {2015}{\natexlab{b}})}\BibitemShut {NoStop}%
\bibitem [{\citenamefont {Xu}\ \emph {et~al.}(2015)\citenamefont {Xu},
  \citenamefont {Belopolski}, \citenamefont {Alidoust}, \citenamefont
  {Neupane}, \citenamefont {Bian}, \citenamefont {Zhang}, \citenamefont
  {Sankar}, \citenamefont {Chang}, \citenamefont {Yuan}, \citenamefont {Lee},
  \citenamefont {Huang}, \citenamefont {Zheng}, \citenamefont {Ma},
  \citenamefont {Sanchez}, \citenamefont {Wang}, \citenamefont {Bansil},
  \citenamefont {Chou}, \citenamefont {Shibayev}, \citenamefont {Lin},
  \citenamefont {Jia},\ and\ \citenamefont {Hasan}}]{Xu613}%
  \BibitemOpen
  \bibfield  {author} {\bibinfo {author} {\bibfnamefont {S.-Y.}\ \bibnamefont
  {Xu}}, \bibinfo {author} {\bibfnamefont {I.}~\bibnamefont {Belopolski}},
  \bibinfo {author} {\bibfnamefont {N.}~\bibnamefont {Alidoust}}, \bibinfo
  {author} {\bibfnamefont {M.}~\bibnamefont {Neupane}}, \bibinfo {author}
  {\bibfnamefont {G.}~\bibnamefont {Bian}}, \bibinfo {author} {\bibfnamefont
  {C.}~\bibnamefont {Zhang}}, \bibinfo {author} {\bibfnamefont
  {R.}~\bibnamefont {Sankar}}, \bibinfo {author} {\bibfnamefont
  {G.}~\bibnamefont {Chang}}, \bibinfo {author} {\bibfnamefont
  {Z.}~\bibnamefont {Yuan}}, \bibinfo {author} {\bibfnamefont {C.-C.}\
  \bibnamefont {Lee}}, \bibinfo {author} {\bibfnamefont {S.-M.}\ \bibnamefont
  {Huang}}, \bibinfo {author} {\bibfnamefont {H.}~\bibnamefont {Zheng}},
  \bibinfo {author} {\bibfnamefont {J.}~\bibnamefont {Ma}}, \bibinfo {author}
  {\bibfnamefont {D.~S.}\ \bibnamefont {Sanchez}}, \bibinfo {author}
  {\bibfnamefont {B.}~\bibnamefont {Wang}}, \bibinfo {author} {\bibfnamefont
  {A.}~\bibnamefont {Bansil}}, \bibinfo {author} {\bibfnamefont
  {F.}~\bibnamefont {Chou}}, \bibinfo {author} {\bibfnamefont {P.~P.}\
  \bibnamefont {Shibayev}}, \bibinfo {author} {\bibfnamefont {H.}~\bibnamefont
  {Lin}}, \bibinfo {author} {\bibfnamefont {S.}~\bibnamefont {Jia}}, \ and\
  \bibinfo {author} {\bibfnamefont {M.~Z.}\ \bibnamefont {Hasan}},\ }\href@noop
  {} {\bibfield  {journal} {\bibinfo  {journal} {Science}\ }\textbf {\bibinfo
  {volume} {349}},\ \bibinfo {pages} {613} (\bibinfo {year}
  {2015})}\BibitemShut {NoStop}%
\bibitem [{\citenamefont {Goerbig}\ \emph {et~al.}(2008)\citenamefont
  {Goerbig}, \citenamefont {Fuchs}, \citenamefont {Montambaux},\ and\
  \citenamefont {Pi\'echon}}]{PhysRevB.78.045415}%
  \BibitemOpen
  \bibfield  {author} {\bibinfo {author} {\bibfnamefont {M.~O.}\ \bibnamefont
  {Goerbig}}, \bibinfo {author} {\bibfnamefont {J.-N.}\ \bibnamefont {Fuchs}},
  \bibinfo {author} {\bibfnamefont {G.}~\bibnamefont {Montambaux}}, \ and\
  \bibinfo {author} {\bibfnamefont {F.}~\bibnamefont {Pi\'echon}},\ }\href@noop
  {} {\bibfield  {journal} {\bibinfo  {journal} {Phys. Rev. B}\ }\textbf
  {\bibinfo {volume} {78}},\ \bibinfo {pages} {045415} (\bibinfo {year}
  {2008})}\BibitemShut {NoStop}%
\bibitem [{\citenamefont {Soluyanov}\ \emph {et~al.}(2015)\citenamefont
  {Soluyanov}, \citenamefont {Gresch}, \citenamefont {Wang}, \citenamefont
  {Wu}, \citenamefont {Troyer}, \citenamefont {Dai},\ and\ \citenamefont
  {Bernevig}}]{Soluyanov2015}%
  \BibitemOpen
  \bibfield  {author} {\bibinfo {author} {\bibfnamefont {A.~A.}\ \bibnamefont
  {Soluyanov}}, \bibinfo {author} {\bibfnamefont {D.}~\bibnamefont {Gresch}},
  \bibinfo {author} {\bibfnamefont {Z.}~\bibnamefont {Wang}}, \bibinfo {author}
  {\bibfnamefont {Q.~Q.-S.}\ \bibnamefont {Wu}}, \bibinfo {author}
  {\bibfnamefont {M.}~\bibnamefont {Troyer}}, \bibinfo {author} {\bibfnamefont
  {X.}~\bibnamefont {Dai}}, \ and\ \bibinfo {author} {\bibfnamefont {B.~A.}\
  \bibnamefont {Bernevig}},\ }\href@noop {} {\bibfield  {journal} {\bibinfo
  {journal} {Nature}\ }\textbf {\bibinfo {volume} {527}},\ \bibinfo {pages}
  {495} (\bibinfo {year} {2015})}\BibitemShut {NoStop}%
\bibitem [{\citenamefont {Wang}\ \emph {et~al.}(2016)\citenamefont {Wang},
  \citenamefont {Gresch}, \citenamefont {Soluyanov}, \citenamefont {Xie},
  \citenamefont {Kushwaha}, \citenamefont {Dai}, \citenamefont {Troyer},
  \citenamefont {Cava},\ and\ \citenamefont
  {Bernevig}}]{PhysRevLett.117.056805}%
  \BibitemOpen
  \bibfield  {author} {\bibinfo {author} {\bibfnamefont {Z.}~\bibnamefont
  {Wang}}, \bibinfo {author} {\bibfnamefont {D.}~\bibnamefont {Gresch}},
  \bibinfo {author} {\bibfnamefont {A.~A.}\ \bibnamefont {Soluyanov}}, \bibinfo
  {author} {\bibfnamefont {W.}~\bibnamefont {Xie}}, \bibinfo {author}
  {\bibfnamefont {S.}~\bibnamefont {Kushwaha}}, \bibinfo {author}
  {\bibfnamefont {X.}~\bibnamefont {Dai}}, \bibinfo {author} {\bibfnamefont
  {M.}~\bibnamefont {Troyer}}, \bibinfo {author} {\bibfnamefont {R.~J.}\
  \bibnamefont {Cava}}, \ and\ \bibinfo {author} {\bibfnamefont {B.~A.}\
  \bibnamefont {Bernevig}},\ }\href@noop {} {\bibfield  {journal} {\bibinfo
  {journal} {Phys. Rev. Lett.}\ }\textbf {\bibinfo {volume} {117}},\ \bibinfo
  {pages} {56805} (\bibinfo {year} {2016})}\BibitemShut {NoStop}%
\bibitem [{\citenamefont {Sun}\ \emph {et~al.}(2015)\citenamefont {Sun},
  \citenamefont {Wu}, \citenamefont {Ali}, \citenamefont {Felser},\ and\
  \citenamefont {Yan}}]{Sun2015}%
  \BibitemOpen
  \bibfield  {author} {\bibinfo {author} {\bibfnamefont {Y.}~\bibnamefont
  {Sun}}, \bibinfo {author} {\bibfnamefont {S.~C.}\ \bibnamefont {Wu}},
  \bibinfo {author} {\bibfnamefont {M.~N.}\ \bibnamefont {Ali}}, \bibinfo
  {author} {\bibfnamefont {C.}~\bibnamefont {Felser}}, \ and\ \bibinfo {author}
  {\bibfnamefont {B.}~\bibnamefont {Yan}},\ }\href@noop {} {\bibfield
  {journal} {\bibinfo  {journal} {Phys. Rev. B}\ }\textbf {\bibinfo {volume}
  {92}},\ \bibinfo {pages} {161107} (\bibinfo {year} {2015})}\BibitemShut
  {NoStop}%
\bibitem [{\citenamefont {Huang}\ \emph {et~al.}(2016)\citenamefont {Huang},
  \citenamefont {McCormick}, \citenamefont {Ochi}, \citenamefont {Zhao},
  \citenamefont {Suzuki}, \citenamefont {Arita}, \citenamefont {Wu},
  \citenamefont {Mou}, \citenamefont {Cao}, \citenamefont {Yan}, \citenamefont
  {Trivedi},\ and\ \citenamefont {Kaminski}}]{Huang2016}%
  \BibitemOpen
  \bibfield  {author} {\bibinfo {author} {\bibfnamefont {L.}~\bibnamefont
  {Huang}}, \bibinfo {author} {\bibfnamefont {T.~M.}\ \bibnamefont
  {McCormick}}, \bibinfo {author} {\bibfnamefont {M.}~\bibnamefont {Ochi}},
  \bibinfo {author} {\bibfnamefont {Z.}~\bibnamefont {Zhao}}, \bibinfo {author}
  {\bibfnamefont {M.-T.}\ \bibnamefont {Suzuki}}, \bibinfo {author}
  {\bibfnamefont {R.}~\bibnamefont {Arita}}, \bibinfo {author} {\bibfnamefont
  {Y.}~\bibnamefont {Wu}}, \bibinfo {author} {\bibfnamefont {D.}~\bibnamefont
  {Mou}}, \bibinfo {author} {\bibfnamefont {H.}~\bibnamefont {Cao}}, \bibinfo
  {author} {\bibfnamefont {J.}~\bibnamefont {Yan}}, \bibinfo {author}
  {\bibfnamefont {N.}~\bibnamefont {Trivedi}}, \ and\ \bibinfo {author}
  {\bibfnamefont {A.}~\bibnamefont {Kaminski}},\ }\href@noop {} {\bibfield
  {journal} {\bibinfo  {journal} {Nat Mater}\ }\textbf {\bibinfo {volume}
  {15}},\ \bibinfo {pages} {1155} (\bibinfo {year} {2016})}\BibitemShut
  {NoStop}%
\bibitem [{\citenamefont {Jiang}\ \emph {et~al.}(2017)\citenamefont {Jiang},
  \citenamefont {Liu}, \citenamefont {Sun}, \citenamefont {Yang}, \citenamefont
  {Rajamathi}, \citenamefont {Qi}, \citenamefont {Yang}, \citenamefont {Chen},
  \citenamefont {Peng}, \citenamefont {Hwang}, \citenamefont {Sun},
  \citenamefont {Mo}, \citenamefont {Vobornik}, \citenamefont {Fujii},
  \citenamefont {Parkin}, \citenamefont {Felser}, \citenamefont {Yan},\ and\
  \citenamefont {Chen}}]{JiangMoTe2}%
  \BibitemOpen
  \bibfield  {author} {\bibinfo {author} {\bibfnamefont {J.}~\bibnamefont
  {Jiang}}, \bibinfo {author} {\bibfnamefont {Z.~K.}\ \bibnamefont {Liu}},
  \bibinfo {author} {\bibfnamefont {Y.}~\bibnamefont {Sun}}, \bibinfo {author}
  {\bibfnamefont {H.~F.}\ \bibnamefont {Yang}}, \bibinfo {author}
  {\bibfnamefont {C.~R.}\ \bibnamefont {Rajamathi}}, \bibinfo {author}
  {\bibfnamefont {Y.~P.}\ \bibnamefont {Qi}}, \bibinfo {author} {\bibfnamefont
  {L.~X.}\ \bibnamefont {Yang}}, \bibinfo {author} {\bibfnamefont
  {C.}~\bibnamefont {Chen}}, \bibinfo {author} {\bibfnamefont {H.}~\bibnamefont
  {Peng}}, \bibinfo {author} {\bibfnamefont {C.-C.}\ \bibnamefont {Hwang}},
  \bibinfo {author} {\bibfnamefont {S.~Z.}\ \bibnamefont {Sun}}, \bibinfo
  {author} {\bibfnamefont {S.-K.}\ \bibnamefont {Mo}}, \bibinfo {author}
  {\bibfnamefont {I.}~\bibnamefont {Vobornik}}, \bibinfo {author}
  {\bibfnamefont {J.}~\bibnamefont {Fujii}}, \bibinfo {author} {\bibfnamefont
  {S.~S.~P.}\ \bibnamefont {Parkin}}, \bibinfo {author} {\bibfnamefont
  {C.}~\bibnamefont {Felser}}, \bibinfo {author} {\bibfnamefont {B.~H.}\
  \bibnamefont {Yan}}, \ and\ \bibinfo {author} {\bibfnamefont {Y.~L.}\
  \bibnamefont {Chen}},\ }\href {\doibase 10.1038/ncomms13973} {\bibfield
  {journal} {\bibinfo  {journal} {Nat Commun}\ }\textbf {\bibinfo {volume}
  {8}},\ \bibinfo {pages} {13973} (\bibinfo {year} {2017})}\BibitemShut
  {NoStop}%
\bibitem [{\citenamefont {Kim}\ \emph {et~al.}(2014)\citenamefont {Kim},
  \citenamefont {Kim},\ and\ \citenamefont {Sasaki}}]{PhysRevB.89.195137}%
  \BibitemOpen
  \bibfield  {author} {\bibinfo {author} {\bibfnamefont {K.-S.}\ \bibnamefont
  {Kim}}, \bibinfo {author} {\bibfnamefont {H.-J.}\ \bibnamefont {Kim}}, \ and\
  \bibinfo {author} {\bibfnamefont {M.}~\bibnamefont {Sasaki}},\ }\href
  {\doibase 10.1103/PhysRevB.89.195137} {\bibfield  {journal} {\bibinfo
  {journal} {Phys. Rev. B}\ }\textbf {\bibinfo {volume} {89}},\ \bibinfo
  {pages} {195137} (\bibinfo {year} {2014})}\BibitemShut {NoStop}%
\bibitem [{\citenamefont {Son}\ and\ \citenamefont
  {Spivak}(2013)}]{PhysRevB.88.104412}%
  \BibitemOpen
  \bibfield  {author} {\bibinfo {author} {\bibfnamefont {D.~T.}\ \bibnamefont
  {Son}}\ and\ \bibinfo {author} {\bibfnamefont {B.~Z.}\ \bibnamefont
  {Spivak}},\ }\href {\doibase 10.1103/PhysRevB.88.104412} {\bibfield
  {journal} {\bibinfo  {journal} {Phys. Rev. B}\ }\textbf {\bibinfo {volume}
  {88}},\ \bibinfo {pages} {104412} (\bibinfo {year} {2013})}\BibitemShut
  {NoStop}%
\bibitem [{\citenamefont {Goswami}\ \emph {et~al.}(2015)\citenamefont
  {Goswami}, \citenamefont {Pixley},\ and\ \citenamefont {{Das
  Sarma}}}]{PhysRevB.92.075205}%
  \BibitemOpen
  \bibfield  {author} {\bibinfo {author} {\bibfnamefont {P.}~\bibnamefont
  {Goswami}}, \bibinfo {author} {\bibfnamefont {J.~H.}\ \bibnamefont {Pixley}},
  \ and\ \bibinfo {author} {\bibfnamefont {S.}~\bibnamefont {{Das Sarma}}},\
  }\href {\doibase 10.1103/PhysRevB.92.075205} {\bibfield  {journal} {\bibinfo
  {journal} {Phys. Rev. B}\ }\textbf {\bibinfo {volume} {92}},\ \bibinfo
  {pages} {75205} (\bibinfo {year} {2015})}\BibitemShut {NoStop}%
\bibitem [{\citenamefont {Gorbar}\ \emph {et~al.}(2014)\citenamefont {Gorbar},
  \citenamefont {Miransky},\ and\ \citenamefont
  {Shovkovy}}]{PhysRevB.89.085126}%
  \BibitemOpen
  \bibfield  {author} {\bibinfo {author} {\bibfnamefont {E.~V.}\ \bibnamefont
  {Gorbar}}, \bibinfo {author} {\bibfnamefont {V.~A.}\ \bibnamefont
  {Miransky}}, \ and\ \bibinfo {author} {\bibfnamefont {I.~A.}\ \bibnamefont
  {Shovkovy}},\ }\href {\doibase 10.1103/PhysRevB.89.085126} {\bibfield
  {journal} {\bibinfo  {journal} {Phys. Rev. B}\ }\textbf {\bibinfo {volume}
  {89}},\ \bibinfo {pages} {85126} (\bibinfo {year} {2014})}\BibitemShut
  {NoStop}%
\bibitem [{\citenamefont {Zhang}\ \emph {et~al.}(2016)\citenamefont {Zhang},
  \citenamefont {Xu}, \citenamefont {Belopolski}, \citenamefont {Yuan},
  \citenamefont {Lin}, \citenamefont {Tong}, \citenamefont {Bian},
  \citenamefont {Alidoust}, \citenamefont {Lee}, \citenamefont {Huang},
  \citenamefont {Chang}, \citenamefont {Chang}, \citenamefont {Hsu},
  \citenamefont {Jeng}, \citenamefont {Neupane}, \citenamefont {Sanchez},
  \citenamefont {Zheng}, \citenamefont {Wang}, \citenamefont {Lin},
  \citenamefont {Zhang}, \citenamefont {Lu}, \citenamefont {Shen},
  \citenamefont {Neupert}, \citenamefont {{Zahid Hasan}},\ and\ \citenamefont
  {Jia}}]{Zhang2016}%
  \BibitemOpen
  \bibfield  {author} {\bibinfo {author} {\bibfnamefont {C.-L.}\ \bibnamefont
  {Zhang}}, \bibinfo {author} {\bibfnamefont {S.-Y.}\ \bibnamefont {Xu}},
  \bibinfo {author} {\bibfnamefont {I.}~\bibnamefont {Belopolski}}, \bibinfo
  {author} {\bibfnamefont {Z.}~\bibnamefont {Yuan}}, \bibinfo {author}
  {\bibfnamefont {Z.}~\bibnamefont {Lin}}, \bibinfo {author} {\bibfnamefont
  {B.}~\bibnamefont {Tong}}, \bibinfo {author} {\bibfnamefont {G.}~\bibnamefont
  {Bian}}, \bibinfo {author} {\bibfnamefont {N.}~\bibnamefont {Alidoust}},
  \bibinfo {author} {\bibfnamefont {C.-C.}\ \bibnamefont {Lee}}, \bibinfo
  {author} {\bibfnamefont {S.-M.}\ \bibnamefont {Huang}}, \bibinfo {author}
  {\bibfnamefont {T.-R.}\ \bibnamefont {Chang}}, \bibinfo {author}
  {\bibfnamefont {G.}~\bibnamefont {Chang}}, \bibinfo {author} {\bibfnamefont
  {C.-H.}\ \bibnamefont {Hsu}}, \bibinfo {author} {\bibfnamefont {H.-T.}\
  \bibnamefont {Jeng}}, \bibinfo {author} {\bibfnamefont {M.}~\bibnamefont
  {Neupane}}, \bibinfo {author} {\bibfnamefont {D.~S.}\ \bibnamefont
  {Sanchez}}, \bibinfo {author} {\bibfnamefont {H.}~\bibnamefont {Zheng}},
  \bibinfo {author} {\bibfnamefont {J.}~\bibnamefont {Wang}}, \bibinfo {author}
  {\bibfnamefont {H.}~\bibnamefont {Lin}}, \bibinfo {author} {\bibfnamefont
  {C.}~\bibnamefont {Zhang}}, \bibinfo {author} {\bibfnamefont {H.-Z.}\
  \bibnamefont {Lu}}, \bibinfo {author} {\bibfnamefont {S.-Q.}\ \bibnamefont
  {Shen}}, \bibinfo {author} {\bibfnamefont {T.}~\bibnamefont {Neupert}},
  \bibinfo {author} {\bibfnamefont {M.}~\bibnamefont {{Zahid Hasan}}}, \ and\
  \bibinfo {author} {\bibfnamefont {S.}~\bibnamefont {Jia}},\ }\href@noop {}
  {\bibfield  {journal} {\bibinfo  {journal} {Nature Communications}\ }\textbf
  {\bibinfo {volume} {7}},\ \bibinfo {pages} {10735} (\bibinfo {year}
  {2016})}\BibitemShut {NoStop}%
\bibitem [{\citenamefont {Hirschberger}\ \emph {et~al.}(2016)\citenamefont
  {Hirschberger}, \citenamefont {Kushwaha}, \citenamefont {Wang}, \citenamefont
  {Gibson}, \citenamefont {Liang}, \citenamefont {Belvin}, \citenamefont
  {Bernevig}, \citenamefont {Cava},\ and\ \citenamefont
  {Ong}}]{Hirschberger2016}%
  \BibitemOpen
  \bibfield  {author} {\bibinfo {author} {\bibfnamefont {M.}~\bibnamefont
  {Hirschberger}}, \bibinfo {author} {\bibfnamefont {S.}~\bibnamefont
  {Kushwaha}}, \bibinfo {author} {\bibfnamefont {Z.}~\bibnamefont {Wang}},
  \bibinfo {author} {\bibfnamefont {Q.}~\bibnamefont {Gibson}}, \bibinfo
  {author} {\bibfnamefont {S.}~\bibnamefont {Liang}}, \bibinfo {author}
  {\bibfnamefont {C.~A.}\ \bibnamefont {Belvin}}, \bibinfo {author}
  {\bibfnamefont {B.~A.}\ \bibnamefont {Bernevig}}, \bibinfo {author}
  {\bibfnamefont {R.~J.}\ \bibnamefont {Cava}}, \ and\ \bibinfo {author}
  {\bibfnamefont {N.~P.}\ \bibnamefont {Ong}},\ }\href@noop {} {\bibfield
  {journal} {\bibinfo  {journal} {Nat Mater}\ }\textbf {\bibinfo {volume}
  {15}},\ \bibinfo {pages} {1161} (\bibinfo {year} {2016})}\BibitemShut
  {NoStop}%
\bibitem [{\citenamefont {Huang}\ \emph {et~al.}(2015)\citenamefont {Huang},
  \citenamefont {Zhao}, \citenamefont {Long}, \citenamefont {Wang},
  \citenamefont {Chen}, \citenamefont {Yang}, \citenamefont {Liang},
  \citenamefont {Xue}, \citenamefont {Weng}, \citenamefont {Fang},
  \citenamefont {Dai},\ and\ \citenamefont {Chen}}]{PhysRevX.5.031023}%
  \BibitemOpen
  \bibfield  {author} {\bibinfo {author} {\bibfnamefont {X.}~\bibnamefont
  {Huang}}, \bibinfo {author} {\bibfnamefont {L.}~\bibnamefont {Zhao}},
  \bibinfo {author} {\bibfnamefont {Y.}~\bibnamefont {Long}}, \bibinfo {author}
  {\bibfnamefont {P.}~\bibnamefont {Wang}}, \bibinfo {author} {\bibfnamefont
  {D.}~\bibnamefont {Chen}}, \bibinfo {author} {\bibfnamefont {Z.}~\bibnamefont
  {Yang}}, \bibinfo {author} {\bibfnamefont {H.}~\bibnamefont {Liang}},
  \bibinfo {author} {\bibfnamefont {M.}~\bibnamefont {Xue}}, \bibinfo {author}
  {\bibfnamefont {H.}~\bibnamefont {Weng}}, \bibinfo {author} {\bibfnamefont
  {Z.}~\bibnamefont {Fang}}, \bibinfo {author} {\bibfnamefont {X.}~\bibnamefont
  {Dai}}, \ and\ \bibinfo {author} {\bibfnamefont {G.}~\bibnamefont {Chen}},\
  }\href@noop {} {\bibfield  {journal} {\bibinfo  {journal} {Phys. Rev. X}\
  }\textbf {\bibinfo {volume} {5}},\ \bibinfo {pages} {31023} (\bibinfo {year}
  {2015})}\BibitemShut {NoStop}%
\bibitem [{\citenamefont {Du}\ \emph {et~al.}(2016)\citenamefont {Du},
  \citenamefont {Wang}, \citenamefont {Chen}, \citenamefont {Mao},
  \citenamefont {Khan}, \citenamefont {Xu}, \citenamefont {Zhou}, \citenamefont
  {Zhang}, \citenamefont {Yang}, \citenamefont {Chen}, \citenamefont {Feng},\
  and\ \citenamefont {Fang}}]{Du59:657406}%
  \BibitemOpen
  \bibfield  {author} {\bibinfo {author} {\bibfnamefont {J.~H.}\ \bibnamefont
  {Du}}, \bibinfo {author} {\bibfnamefont {H.~D.}\ \bibnamefont {Wang}},
  \bibinfo {author} {\bibfnamefont {Q.}~\bibnamefont {Chen}}, \bibinfo {author}
  {\bibfnamefont {Q.~H.}\ \bibnamefont {Mao}}, \bibinfo {author} {\bibfnamefont
  {R.}~\bibnamefont {Khan}}, \bibinfo {author} {\bibfnamefont {B.~J.}\
  \bibnamefont {Xu}}, \bibinfo {author} {\bibfnamefont {Y.~X.}\ \bibnamefont
  {Zhou}}, \bibinfo {author} {\bibfnamefont {Y.~N.}\ \bibnamefont {Zhang}},
  \bibinfo {author} {\bibfnamefont {J.~H.}\ \bibnamefont {Yang}}, \bibinfo
  {author} {\bibfnamefont {B.}~\bibnamefont {Chen}}, \bibinfo {author}
  {\bibfnamefont {C.~M.}\ \bibnamefont {Feng}}, \ and\ \bibinfo {author}
  {\bibfnamefont {M.~H.}\ \bibnamefont {Fang}},\ }\href {\doibase
  10.1007/s11433-016-5798-4} {\bibfield  {journal} {\bibinfo  {journal}
  {Science China Physics, Mechanics and Astronomy}\ }\textbf {\bibinfo {volume}
  {59}},\ \bibinfo {pages} {657406} (\bibinfo {year} {2016})}\BibitemShut
  {NoStop}%
\bibitem [{\citenamefont {Watzman}\ \emph {et~al.}(2017)\citenamefont
  {Watzman}, \citenamefont {McCormick}, \citenamefont {Sekhar}, \citenamefont
  {Wu}, \citenamefont {Sun}, \citenamefont {Prakash}, \citenamefont {Felser},
  \citenamefont {Trivedi},\ and\ \citenamefont {Heremans}}]{nbpNernst}%
  \BibitemOpen
  \bibfield  {author} {\bibinfo {author} {\bibfnamefont {S.}~\bibnamefont
  {Watzman}}, \bibinfo {author} {\bibfnamefont {T.~M.}\ \bibnamefont
  {McCormick}}, \bibinfo {author} {\bibfnamefont {C.}~\bibnamefont {Sekhar}},
  \bibinfo {author} {\bibfnamefont {S.-C.}\ \bibnamefont {Wu}}, \bibinfo
  {author} {\bibfnamefont {Y.}~\bibnamefont {Sun}}, \bibinfo {author}
  {\bibfnamefont {A.}~\bibnamefont {Prakash}}, \bibinfo {author} {\bibfnamefont
  {C.}~\bibnamefont {Felser}}, \bibinfo {author} {\bibfnamefont
  {N.}~\bibnamefont {Trivedi}}, \ and\ \bibinfo {author} {\bibfnamefont
  {J.~P.}\ \bibnamefont {Heremans}},\ }\href@noop {} {} (\bibinfo {year}
  {2017}),\ \Eprint {http://arxiv.org/abs/arXiv:1703.04700} {arXiv:1703.04700}
  \BibitemShut {NoStop}%
\bibitem [{\citenamefont {Lundgren}\ \emph {et~al.}(2014)\citenamefont
  {Lundgren}, \citenamefont {Laurell},\ and\ \citenamefont
  {Fiete}}]{fieteThermoelec}%
  \BibitemOpen
  \bibfield  {author} {\bibinfo {author} {\bibfnamefont {R.}~\bibnamefont
  {Lundgren}}, \bibinfo {author} {\bibfnamefont {P.}~\bibnamefont {Laurell}}, \
  and\ \bibinfo {author} {\bibfnamefont {G.~A.}\ \bibnamefont {Fiete}},\
  }\href@noop {} {\bibfield  {journal} {\bibinfo  {journal} {Phys. Rev. B}\
  }\textbf {\bibinfo {volume} {90}},\ \bibinfo {pages} {165115} (\bibinfo
  {year} {2014})}\BibitemShut {NoStop}%
\bibitem [{\citenamefont {Sharma}\ \emph {et~al.}(2016)\citenamefont {Sharma},
  \citenamefont {Goswami},\ and\ \citenamefont {Tewari}}]{PhysRevB.93.035116}%
  \BibitemOpen
  \bibfield  {author} {\bibinfo {author} {\bibfnamefont {G.}~\bibnamefont
  {Sharma}}, \bibinfo {author} {\bibfnamefont {P.}~\bibnamefont {Goswami}}, \
  and\ \bibinfo {author} {\bibfnamefont {S.}~\bibnamefont {Tewari}},\
  }\href@noop {} {\bibfield  {journal} {\bibinfo  {journal} {Phys. Rev. B}\
  }\textbf {\bibinfo {volume} {93}},\ \bibinfo {pages} {035116} (\bibinfo
  {year} {2016})}\BibitemShut {NoStop}%
\bibitem [{\citenamefont {McCormick}\ \emph {et~al.}(2018)\citenamefont
  {McCormick}, \citenamefont {Watzman}, \citenamefont {Heremans},\ and\
  \citenamefont {Trivedi}}]{arcTherm}%
  \BibitemOpen
  \bibfield  {author} {\bibinfo {author} {\bibfnamefont {T.~M.}\ \bibnamefont
  {McCormick}}, \bibinfo {author} {\bibfnamefont {S.~J.}\ \bibnamefont
  {Watzman}}, \bibinfo {author} {\bibfnamefont {J.~P.}\ \bibnamefont
  {Heremans}}, \ and\ \bibinfo {author} {\bibfnamefont {N.}~\bibnamefont
  {Trivedi}},\ }\href@noop {} {\bibfield  {journal} {\bibinfo  {journal} {Phys.
  Rev. B}\ }\textbf {\bibinfo {volume} {97}},\ \bibinfo {pages} {195152}
  (\bibinfo {year} {2018})}\BibitemShut {NoStop}%
\bibitem [{\citenamefont {McCormick}\ \emph
  {et~al.}(2017{\natexlab{a}})\citenamefont {McCormick}, \citenamefont
  {McKay},\ and\ \citenamefont {Trivedi}}]{mccormickmckay}%
  \BibitemOpen
  \bibfield  {author} {\bibinfo {author} {\bibfnamefont {T.~M.}\ \bibnamefont
  {McCormick}}, \bibinfo {author} {\bibfnamefont {R.~C.}\ \bibnamefont
  {McKay}}, \ and\ \bibinfo {author} {\bibfnamefont {N.}~\bibnamefont
  {Trivedi}},\ }\href@noop {} {\bibfield  {journal} {\bibinfo  {journal} {Phys.
  Rev. B}\ }\textbf {\bibinfo {volume} {96}},\ \bibinfo {pages} {235116}
  (\bibinfo {year} {2017}{\natexlab{a}})}\BibitemShut {NoStop}%
\bibitem [{\citenamefont {Wu}\ \emph {et~al.}(2015)\citenamefont {Wu},
  \citenamefont {Jo}, \citenamefont {Ochi}, \citenamefont {Huang},
  \citenamefont {Mou}, \citenamefont {Bud'ko}, \citenamefont {Canfield},
  \citenamefont {Trivedi}, \citenamefont {Arita},\ and\ \citenamefont
  {Kaminski}}]{wte2liftrans}%
  \BibitemOpen
  \bibfield  {author} {\bibinfo {author} {\bibfnamefont {Y.}~\bibnamefont
  {Wu}}, \bibinfo {author} {\bibfnamefont {N.~H.}\ \bibnamefont {Jo}}, \bibinfo
  {author} {\bibfnamefont {M.}~\bibnamefont {Ochi}}, \bibinfo {author}
  {\bibfnamefont {L.}~\bibnamefont {Huang}}, \bibinfo {author} {\bibfnamefont
  {D.}~\bibnamefont {Mou}}, \bibinfo {author} {\bibfnamefont {S.~L.}\
  \bibnamefont {Bud'ko}}, \bibinfo {author} {\bibfnamefont {P.~C.}\
  \bibnamefont {Canfield}}, \bibinfo {author} {\bibfnamefont {N.}~\bibnamefont
  {Trivedi}}, \bibinfo {author} {\bibfnamefont {R.}~\bibnamefont {Arita}}, \
  and\ \bibinfo {author} {\bibfnamefont {A.}~\bibnamefont {Kaminski}},\ }\href
  {\doibase 10.1103/PhysRevLett.115.166602} {\bibfield  {journal} {\bibinfo
  {journal} {Phys. Rev. Lett.}\ }\textbf {\bibinfo {volume} {115}},\ \bibinfo
  {pages} {166602} (\bibinfo {year} {2015})}\BibitemShut {NoStop}%
\bibitem [{\citenamefont {Ferreiros}\ \emph {et~al.}(2017)\citenamefont
  {Ferreiros}, \citenamefont {Zyuzin},\ and\ \citenamefont
  {Bardarson}}]{NernstFerreiros}%
  \BibitemOpen
  \bibfield  {author} {\bibinfo {author} {\bibfnamefont {Y.}~\bibnamefont
  {Ferreiros}}, \bibinfo {author} {\bibfnamefont {A.~A.}\ \bibnamefont
  {Zyuzin}}, \ and\ \bibinfo {author} {\bibfnamefont {J.~H.}\ \bibnamefont
  {Bardarson}},\ }\href {\doibase 10.1103/PhysRevB.96.115202} {\bibfield
  {journal} {\bibinfo  {journal} {Phys. Rev. B}\ }\textbf {\bibinfo {volume}
  {96}},\ \bibinfo {pages} {115202} (\bibinfo {year} {2017})}\BibitemShut
  {NoStop}%
\bibitem [{\citenamefont {McCormick}\ \emph
  {et~al.}(2017{\natexlab{b}})\citenamefont {McCormick}, \citenamefont
  {Kimchi},\ and\ \citenamefont {Trivedi}}]{mkt}%
  \BibitemOpen
  \bibfield  {author} {\bibinfo {author} {\bibfnamefont {T.~M.}\ \bibnamefont
  {McCormick}}, \bibinfo {author} {\bibfnamefont {I.}~\bibnamefont {Kimchi}}, \
  and\ \bibinfo {author} {\bibfnamefont {N.}~\bibnamefont {Trivedi}},\
  }\href@noop {} {\bibfield  {journal} {\bibinfo  {journal} {Phys. Rev. B}\
  }\textbf {\bibinfo {volume} {95}},\ \bibinfo {pages} {075133} (\bibinfo
  {year} {2017}{\natexlab{b}})}\BibitemShut {NoStop}%
\bibitem [{\citenamefont {Bernevig}\ and\ \citenamefont
  {Hughes}(2013)}]{bernBook}%
  \BibitemOpen
  \bibfield  {author} {\bibinfo {author} {\bibfnamefont {A.}~\bibnamefont
  {Bernevig}}\ and\ \bibinfo {author} {\bibfnamefont {T.~L.}\ \bibnamefont
  {Hughes}},\ }\href@noop {} {\emph {\bibinfo {title} {Topological Insulators
  and Topological Superconductors}}}\ (\bibinfo  {publisher} {Princeton},\
  \bibinfo {year} {2013})\BibitemShut {NoStop}%
\bibitem [{\citenamefont {Harman}\ and\ \citenamefont
  {Honig}(1967)}]{harmHonig}%
  \BibitemOpen
  \bibfield  {author} {\bibinfo {author} {\bibfnamefont {T.~C.}\ \bibnamefont
  {Harman}}\ and\ \bibinfo {author} {\bibfnamefont {J.~M.}\ \bibnamefont
  {Honig}},\ }\href@noop {} {\emph {\bibinfo {title} {Thermoelectric and
  Thermomagnetic Effects and Applications}}}\ (\bibinfo  {publisher}
  {Macgraw-Hill},\ \bibinfo {address} {New York},\ \bibinfo {year}
  {1967})\BibitemShut {NoStop}%
\bibitem [{\citenamefont {Ashcroft}\ and\ \citenamefont
  {Mermin}(1976)}]{ashcroftbook}%
  \BibitemOpen
  \bibfield  {author} {\bibinfo {author} {\bibfnamefont {N.}~\bibnamefont
  {Ashcroft}}\ and\ \bibinfo {author} {\bibfnamefont {N.}~\bibnamefont
  {Mermin}},\ }\href@noop {} {\bibfield  {journal} {\bibinfo  {journal}
  {Saunder College, Philadelphia}\ } (\bibinfo {year} {1976})}\BibitemShut
  {NoStop}%
\bibitem [{\citenamefont {Chen}\ and\ \citenamefont
  {Fiete}(2016)}]{doubleWeyl}%
  \BibitemOpen
  \bibfield  {author} {\bibinfo {author} {\bibfnamefont {Q.}~\bibnamefont
  {Chen}}\ and\ \bibinfo {author} {\bibfnamefont {G.~A.}\ \bibnamefont
  {Fiete}},\ }\href@noop {} {\bibfield  {journal} {\bibinfo  {journal} {Phys.
  Rev. B}\ }\textbf {\bibinfo {volume} {93}},\ \bibinfo {pages} {155125}
  (\bibinfo {year} {2016})}\BibitemShut {NoStop}%
\bibitem [{\citenamefont {Xiao}\ \emph {et~al.}(2006)\citenamefont {Xiao},
  \citenamefont {Yao}, \citenamefont {Fang},\ and\ \citenamefont
  {Niu}}]{anomNernstNiu}%
  \BibitemOpen
  \bibfield  {author} {\bibinfo {author} {\bibfnamefont {D.}~\bibnamefont
  {Xiao}}, \bibinfo {author} {\bibfnamefont {Y.}~\bibnamefont {Yao}}, \bibinfo
  {author} {\bibfnamefont {Z.}~\bibnamefont {Fang}}, \ and\ \bibinfo {author}
  {\bibfnamefont {Q.}~\bibnamefont {Niu}},\ }\href {\doibase
  10.1103/PhysRevLett.97.026603} {\bibfield  {journal} {\bibinfo  {journal}
  {Phys. Rev. Lett.}\ }\textbf {\bibinfo {volume} {97}},\ \bibinfo {pages}
  {026603} (\bibinfo {year} {2006})}\BibitemShut {NoStop}%
\bibitem [{\citenamefont {Qin}\ \emph {et~al.}(2011)\citenamefont {Qin},
  \citenamefont {Niu},\ and\ \citenamefont {Shi}}]{thermhall3}%
  \BibitemOpen
  \bibfield  {author} {\bibinfo {author} {\bibfnamefont {T.}~\bibnamefont
  {Qin}}, \bibinfo {author} {\bibfnamefont {Q.}~\bibnamefont {Niu}}, \ and\
  \bibinfo {author} {\bibfnamefont {J.}~\bibnamefont {Shi}},\ }\href@noop {}
  {\bibfield  {journal} {\bibinfo  {journal} {Phys. Rev. Lett.}\ }\textbf
  {\bibinfo {volume} {107}},\ \bibinfo {pages} {236601} (\bibinfo {year}
  {2011})}\BibitemShut {NoStop}%
\bibitem [{\citenamefont {Lauer}(1974)}]{lauer}%
  \BibitemOpen
  \bibfield  {author} {\bibinfo {author} {\bibfnamefont {R.~B.}\ \bibnamefont
  {Lauer}},\ }\href@noop {} {\bibfield  {journal} {\bibinfo  {journal} {J.
  Appl. Phys.}\ }\textbf {\bibinfo {volume} {45}},\ \bibinfo {pages} {1794}
  (\bibinfo {year} {1974})}\BibitemShut {NoStop}%
\end{thebibliography}%

\end{document}